\documentclass[
preprint,
superscriptaddress,
 amsmath,amssymb,
 aps,
prb,
floatfix,
]{revtex4-1}

\usepackage{graphicx}
\usepackage{dcolumn}
\usepackage{bm}

\usepackage{xcolor}
\newcommand{\rev}[1]{\textcolor{red}{#1}}

\usepackage{ulem}

\begin{document}

\preprint{APS/123-QED}

\title{Angular dependence and absorption properties of the anapole mode of Si nano-disks}

\author{L. Fornasari}
\affiliation{Plasmore s.r.l., viale Vittorio Emanuele II, 4, I-27100, Pavia (Italy)}

\author{M. Passoni}
\author{F. Marabelli}
\email[]{franco.marabelli@unipv.it}
\affiliation{Physics Department, University of Pavia, via Bassi,6, I-27100, Pavia (Italy)}

\author{Y. Chen}
\affiliation{Department of Electrical and Computer Engineering \& Photonics Center, Boston University, 8 Saint Mary's Street, Boston, MA 02215, USA}

\author{Y. Wang}
\affiliation{Department of Electrical and Computer Engineering \& Photonics Center, Boston University, 8 Saint Mary's Street, Boston, MA 02215, USA}

\author{L. Dal Negro}
\affiliation{Department of Electrical and Computer Engineering \& Photonics Center, Boston University, 8 Saint Mary's Street, Boston, MA 02215, USA}
\affiliation{Boston University, Division of Material Science and Engineering, 15 Saint Mary’s Street, Boston, Massachusetts, 02446, USA}
\affiliation{Boston University, Department of Physics, 590 Commonwealth Avenue, Boston, Massachusetts, 02215, USA}

\date{\today}

\begin{abstract}
The polarization- and angle- resolved optical response of the anapole mode in silicon nano-disks array have been experimentally and theoretically investigated. The good agreement between measured data and simulations yields to a consistent description of the anapole mode behavior that exhibits different features for TE or TM polarization excitation. Scattering matrix calculation allows us to disentangle scattered and diffused light contributions and to provide a quantitative estimation of the absorbance enhancement associated to 2D excitation of the anapole mode. We performed the multipolar decomposition of the far-field scattered radiation for both TE and TM polarizations and unambiguously identified the anapole resonant condition in excellent agreement with the experimental results over a large range of incident angles. Our findings demonstrate the controlled excitation of electromagnetic anapole modes in engineered arrays of silicon nano-disks for the development of optical nanostructures with enhanced light-matter interaction.   
\end{abstract}

\maketitle


\section{Introduction}

Since the development of optics in terms of electromagnetic interaction, scattering of light by small particles has been the subject of a number of studies. This interest raised enormously during the last decade in connection with the rapid advancements of nano-optical technologies. Light-matter interactions at the nanoscale offer novel possibilities to manipulate and shape radiation using engineered nano-sized materials and devices. Among the many reported nano-optical effects, the intriguing anapole behavior lately attracted significant attention \cite{AdvOptMat2019} \sout{,} thanks to its ability to completely suppress radiation and scattering \cite{miroshnichenko2015nonradiating}. In dielectric nano-particles, the anapole condition has been associated to the vanishing of the spherical electric dipole mode due to the almost exact cancellation of the Cartesian electric and toroidal radiation diagrams \cite{Wei:16}.
The effect has been observed in dielectric spherical nanoparticles, as well as in disk\rev{-}shaped ones \cite{Svyakhovskiy:19}. Moreover, the capability of fabricating even more complex nanostructures, including also metallic components, increased the possibilities of observing this phenomenon \cite{papasimakis_electromagnetic_2016}. In very recent time many theoretical and numerical studies have been published on this subject \cite{ospanova_multipolar_2018,gongora_anapole_2017,wu_strong_2020,savinov_optical_2019}. However, few experimental results have been reported in the optical regime so far \cite{yang_nonradiating_2019,tian_active_2019} and fundamental questions are still a matter of debate. It is generally believed that the anapole behavior originates from a peculiar distribution of localized electromagnetic field inside nano-particles, but it is not clear whether this corresponds to a real excitation mode of the system, how it can be effectively excited, and to what degree it affects the absorption properties of nanostructures \cite{monticone2019can}. For instance, a recent calculation shows the possibility of obtaining extreme values of electric field enhancement exploiting the anapole condition \cite{acsphotonics2018}. Another work experimentally demonstrates the enhancement of light-matter interaction at the anapole conditions through Raman scattering experiments \cite{nanoscaleAlex}. However, it is important to realize that the anapole condition has been mainly investigated under normal incidence, although non-trivial effects could be observed for different excitation angles \cite{paniagua2016generalized}.

In this work, by combining numerical simulations and experiments we investigate  the angular dispersion and polarization dependence of the anapoles excited in arrays of Si nano-disks with different geometrical parameters. Moreover, we estimate the field enhancement associated to the anapole condition and report on the corresponding absorption enhancement. 

Square arrays of Si-disks were fabricated on silica substrates using standardized electron beam lithography and etching processes described elsewhere \cite{lawrence2012aperiodic, gorsky2018directional,gorsky2019engineered,britton2019indium}. Three series of samples have been prepared with disk height = 60 nm and diameters of 300, 380 and 430 nm (array 1, 2 and 3, respectively). The lattice pitch is 2 $\mu$m for all samples.
Angle- and polarization-resolved reflectance (R) and transmittance (T) measurements have been performed in the spectral range 500-1100 nm using a Fourier Transform spectrometer with a silicon photodiode detector coupled to a custom-made micro-reflectometer \cite{giudicatti_interaction_2012}. 
Measurements have been performed from normal incidence up to $70^\circ$, with beam divergence below $1^\circ$. A light spot with a diameter of about 100 $\mu m$ has been used to excite the patterned area of the sample.We used a Glan-Taylor calcite polarizer to select the excitation with electric field transverse (TE) or parallel (TM) to the plane of incidence. A reference measurement has been collected on an unpatterned area. The proportion of absorbed/diffused light with respect to the incident light has been estimated from the difference 1-R-T. A small contribution (less than $1\%$) originating from the scattering in the substrate has been subtracted. 

\begin{figure}[t]
        {\includegraphics[width=0.31\textwidth]{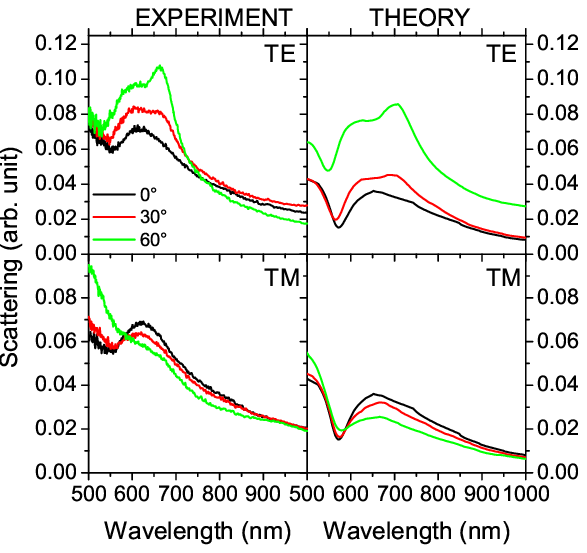}}
        {\includegraphics[width=0.31\textwidth]{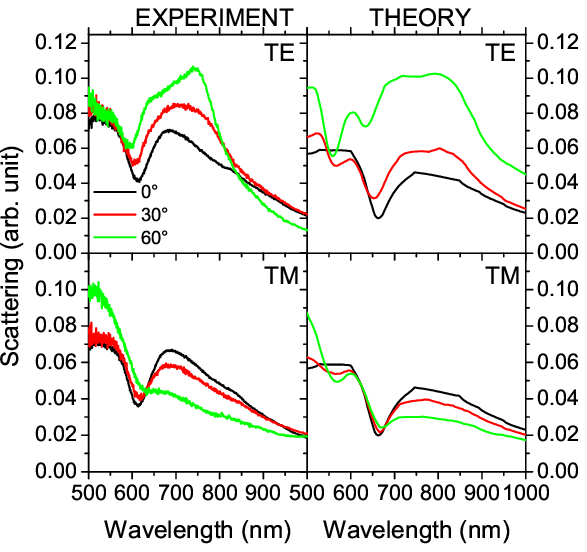}}
        {\includegraphics[width=0.31\textwidth]{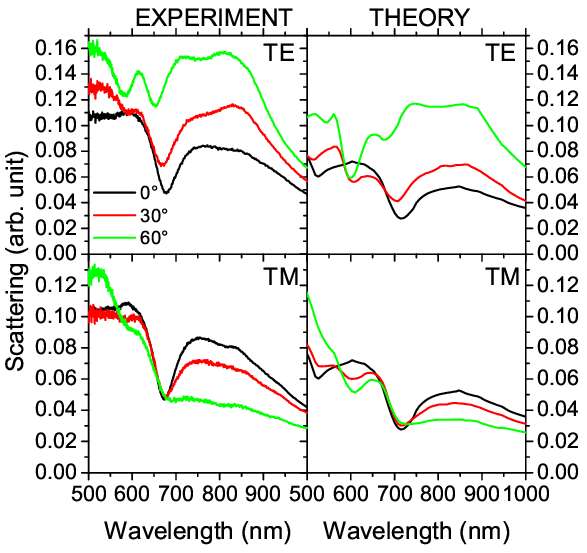}}
    \caption{Sum of the absorbance and diffusion contribution for Array 1 (left), Array 2 (center) and Array 3 (right), respectively. The left side plot shows the experiment data for TE and TM polarized light. In the right side panel the corresponding simulation results are presented. In each plot the spectra for $0^\circ$ (black line), $30^\circ$ (red line) and $60^\circ$ (green line) incidence angles are shown.}\label{fig:Figura1.eps}
\end{figure}

Numerical simulations of the same structures have been performed with an in-house python implementation of the scattering matrix method, using the formulation first proposed by Lifeng Li for applications to crossed gratings \cite{li_new_1997}. This method allows us to unambiguously differentiate the various scattering channels. In particular, we focused on the transmittance and reflectance at the $0^{th}$ diffractive order, the total scattered power considering all diffraction orders, as well as the total absorption in the silicon-disks. The numerical convergence has been accurately evaluated (see Supplementary Information for details). Simulation results were obtained using 961 plane waves with a square truncation scheme that we estimated to yield a good balance between accuracy and computational time across the investigated spectrum. 
In our simulations we considered a square lattice of silicon pillars on a semi-infinite silica substrate. As a result, the  reflection contribution (almost constant, $\sim$ 4\% ) form the backside of the substrate is not considered here. This missing contribution leads to smaller calculated absorption/diffusion values with respect to the experimental values.
Pillar radius has been slightly reduced (-10 nm) with respect to the nominal values in order to take into account the effect of silicon oxidation and match the positions of the spectrum minima obtained by the experiment.

Fig. \ref{fig:Figura1.eps} shows the comparison between the experimental and the calculated results of the sum of diffusion and absorption for the three samples.
Despite the discrepancy in the absolute value, the behavior of the experimental data is quite well-reproduced by the simulations. In particular, both the angle- and the pillar size- dependence of the absorbance/diffusion spectra is well-described. The minimum of the scattering spectrum corresponds to the anapole mode, which appears at 550 nm for array 1 at $0^\circ$ incidence, while at the same angle it is red shifted up to about 600 nm  and 700 nm for arrays 2 and 3, respectively. The measured intensities were found to increase for increasing angles for the TE polarization while to decrease for the TM polarization. Moreover, the minimum of the scattering  blue-shifts for increasing angles in the TE mode while it remains at the same wavelength in the TM mode.
Another peculiar feature is that the diffusion maximum around 600 nm at $0^\circ$ for array 1 (680 nm for array 2 and 770 nm for array 3), splits into two maxima when crossing $60^\circ$ incidence. No similar effect is observed for TM polarization mode. 

In order to deepen our understanding on the sensitivity of anapole modes to the incident polarization and incidence angle, we performed a three-dimensional Finite Element Method (FEM) simulations and multipolar decomposition of the scattered radiation under various incidence and polarization conditions. Indeed the excitation of non-radiating anapole modes has been largely investigated under normal incidence conditions. In the past\rev{,} under such a condition it was found that the far-field scattering intensity of dielectric nanoparticles is strongly reduced and consequently the absorption efficiency is enhanced due to destructive interference of the electric dipole modes and the toroidal dipole modes \cite{papasimakis_electromagnetic_2016,kaelberer_toroidal_2010}. Si nano-disks at normal incidence has been investigated before using the electromagnetic multipolar decomposition method \cite{kaelberer_toroidal_2010,miroshnichenko_nonradiating_2015,wang_engineering_2016}.

\begin{table*}[t]
  \begin{center}
    \caption{Far field Scattering Power Expressions}
    \label{tab:table1}
    \begin{tabular}{l|c|r} 
\hline
      \textbf{Multipoles} & \textbf{Expressions} & \textbf{Far-field scattering power}\\
\hline
      Electric Dipole (p) & $p=\frac{1}{i\omega}\int J(r) d^3r$ & $I_p=\frac{2\omega^4}{3c^3}|p|^2$ \\   
      Magnetic Dipole (m) & $m=\frac{1}{ic}\int r\times J(r) d^3r$ & $I_p=\frac{2\omega^4}{3c^3}|m|^2$ \\
      Toroidal Dipole (T) & $T=\frac{1}{10c}\int \{[r \cdot  J(r)]r-2r^2J(r)\} d^3r$ & $I_p=\frac{2\omega^6}{3c^5}|T|^2$ \\
      Electric Quadrupole (Q$_e$) & $Q_{\alpha \beta}=\frac{1}{2i \omega}\int \{r_\alpha  J_\beta (r)+r_\beta  J_\alpha (r)-\frac{2}{3} [r\cdot J(r)]\delta_{\alpha \beta} \} d^3r$ & $I_Q^e=\frac{\omega^6}{5c^5} \sum_{\alpha \beta} |Q_{\alpha \beta}|^2$ \\
Magnetic Quadrupole (Q$_m$) & $M_{\alpha \beta}=\frac{1}{3c}\int \{[r \times  J(r)]_\alpha r_\beta  +[r \times J(r)]_\beta r_\alpha (r) \} d^3r$ & $I_Q^m=\frac{\omega^6}{20c^5} \sum_{\alpha \beta} |M_{\alpha \beta}|^2$ \\
\hline
    \end{tabular}
  \end{center}
\end{table*}

\begin{figure}[t!]
\includegraphics[width=0.45\textwidth]{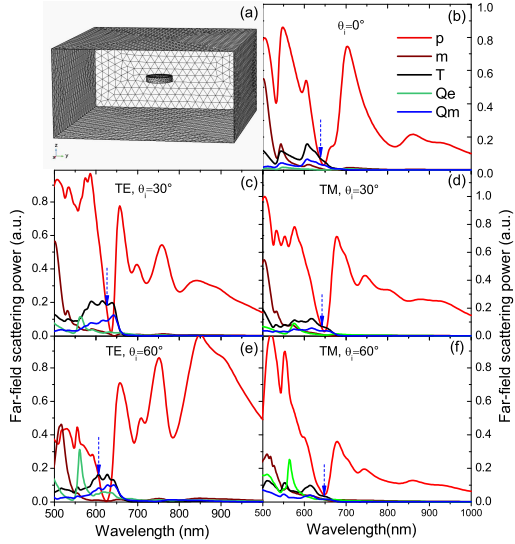}
\caption{(a) Schematic drawing of the model used for simulation. The multipolar decomposition of the five multipole moments: electric dipole (p, red line), magnetic dipole (m, wine line), toroidal dipole (T, black line), electric quadrupole (Qe, green line), magnetic quadrupole (Qm, purple line) for the nano-disk array is calculated for different incident angles and polarization states: (b) TE and TM at normal incidence; (c) TE at $30^\circ$; (d) TM at $30^\circ$, (e) TE at $60^\circ$; (f) TM at $60^\circ$. The anapole mode position where the intensities of the electric dipole and toroidal dipole are equal is indicated by the blue arrow.\label{fig:Figura2}}
\end{figure}

We modeled the samples as a Si disk at the center of a unit cell  (cell side $d=2 \mu m$), as shown in Fig. \ref{fig:Figura2}(a). Periodic boundary conditions were imposed on the four lateral sides of the unit cell. Moreover, we implemented the scattering boundary conditions for the top and bottom surfaces of the unit cell and we excited the array using plane wave excitation from the top surface. The minimum mesh size used in the simulation was $8 nm$. We defined the angle between the incoming wave vector and the z-axis as $\theta_i$. We then numerically obtained the induced current distribution inside the nano-disks and calculated the far-field scattering power for each electromagnetic multipole by volume integration using the expressions tabulated in Table 1 \cite{wang_engineering_2016}. The indices $\alpha$ and $\beta$ represent the Cartesian axes x, y and z in the electric and magnetic quadrupoles. In the expression of the electric quadrupole, the $\delta_{\alpha\beta}$ is the Kronecker delta.

\begin{figure}[b]
	\centering
		\includegraphics[width=0.5\textwidth]{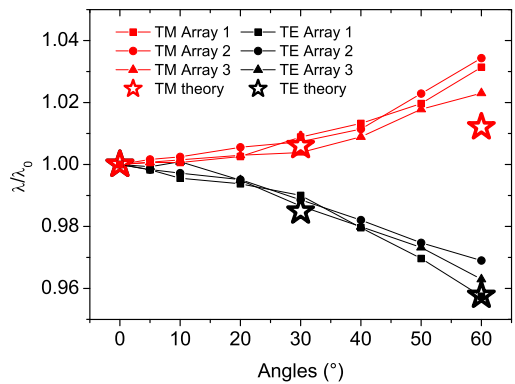}
	\caption{Dependence of the anapole position on the incidence angles. To compare the shift between theory and experiment and among different samples, all the data have been normalized on the anapole wavelength at normal incidence. }
	\label{fig:Figure3}
\end{figure}

In Fig. \ref{fig:Figura2}(b) we show the calculated spectra of the electromagnetic multipoles that contribute to the far-field scattering radiation of an array of Si nano-disks with diameter d = 300 nm and height h = 60 nm, excited at normal incidence. At the positions indicated by the dashed blue arrows the interference between the electric dipole and the toroidal electric mode leads to the almost total cancellation of the far-field scattered radiation. In Figs. \ref{fig:Figura2}(c)-(f) we show the multipolar decomposition obtained by varying the incident angle of illumination from normal incidence to oblique incidence (up to $60^\circ$). We found that the anapole mode under TE excitation shifts to shorter wavelengths (from 638 nm to 605 nm) in qualitative good agreement with the experimental measurements. On the other hand, the spectral position of the anapole mode remains almost exactly at the same wavelength under TM excitation, also in agreement with our measured data. 
A comparison between theory and experiment of the normalized shifts of the anapole mode positions for all the fabricated arrays is reported in Fig. \ref{fig:Figure3}, where we can observe a very good agreement between the simulation and experimental results.

\begin{figure}[t]
    \centering
        \includegraphics[width=0.5\textwidth]{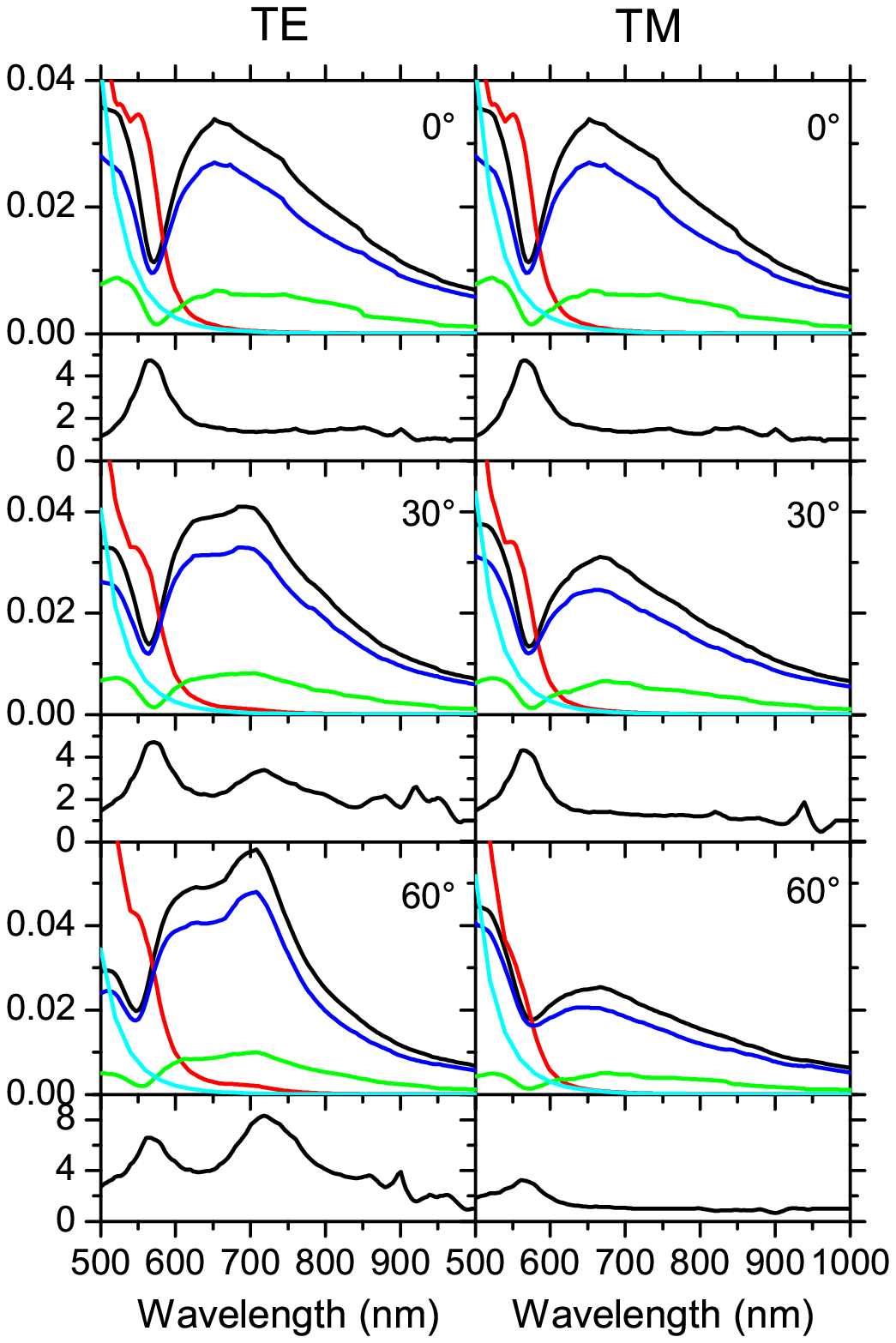}
    \caption{Array 1 scattering and absorption spectrum (black line), obtained with scattering matrix calculation, has been disentangled in the different contributions: the reflected (green line) and transmitted (blue line) contribution of the diffuse and the absorption (red line). The absorbance of a layer of an equivalent amount of Silicon (calculated with an incoherent model) is also reported (light blue line). Both the shown absorbance data have been multiplied by a factor 10. The black lines in the small panels highlight the ratio between the two absorbance. Data for $0^\circ$, $30^\circ$ and $60^\circ$ angles of incidence with TE and TM light polarization have been reported.}\label{fig:Figure4}
\end{figure}
 
The scattering matrix calculations allow us to distinguish between absorbed and scattered/diffused light. In Fig.\ref{fig:Figure4} we plot the different component of scattering/absorption at at different angles of incidence. Similar results are obtained for array 2 and 3 and are reported in Supporting Information. The scattered light can be further decomposed into the part reflected back from the surface (the upward component) and the one transmitted and scattered towards the back side (the downward component). It is evident that scattering dominates the process and determines the spectral features observed experimentally above. A direct correlation of these features can be found with the electric dipole contribution spectrum as obtained by FEM simulations.
It is anyways interesting to look at the absorption spectrum as well. Although the intensity is very low, definite spectral features can be observed. In particular, an absorption peak appears in correspondence due to the anapole mode excitation corresponding to the minimum of the scattered radiation. In order to quantify this effect, we perform a comparison of the absorbance spectrum with the one of an equivalent amount of silicon. Considering the given pillar dimensions (diameter, height) and density (pitch), one can verify that the silicon amount of the array 1 is equivalent to a uniform Si layer of 1 nm thickness (1.7 nm for array 2 and 2.2 nm for array 3). The absorbance of such an equivalent Si layer is calculated using an incoherent model to account for multiple reflections, in order to avoid interference fringes. We notice that just slightly smaller values for the equivalent thickness can be obtained by considering absorbance of a film at the nominal value (60 nm), rescaled by surface filling fraction of the Si nano-disks). Results of such a calculation have been added to the plots in Fig. \ref{fig:Figure4} where all absorbance data are multiplied by a factor 10 to improve the visibility. The spectral dependence of the ratio between the absorbance calculated for the anapole structure and the equivalent Si film absorbance is shown by the back curves at the bottom of Fig. 4. We remark that at the wavelength where anapole occurs this ratio is almost 5, demonstrating the importance of this effect for absorption enhancement applications. However, one may notice that besides the peak at the anapole wavelength, in the TE polarization at the largest angles of incidence, additional maxima emerge at longer wavelengths. In the case of array 1 a maximum appears just above 700 nm with growing incident angles. The absorbance of array 2 presents two additional maxima at around 550 nm and 850 nm while array 3 shows maxima at 600 nm and 900 nm at the highest incident angles. 
These behaviors reflect the angle dependence of the electric dipole scattering which is related to the the strongly anisotropic electric field distribution in the studied system.

In conclusion, by combining accurate numerical simulations with angle- and polarization- dependent scattering measurements, we demonstrate significant absorption enhancement in fabricated Si nano-disk arrays under anapole excitation conditions. Furthermore, we provide a systematic characterization of the anapole dependence on the excitation angles and polarization conditions. These findings motivate the application of anapole-driven electromagnetic responses to arrays of dieletric nano-disks for the engineering of more efficient active devices, such as nano-lasers and detectors on the Si platform.
\begin{acknowledgments}
L.D.N would like to thank the partial support of the Army Research Laboratory under Cooperative Agreement Number W911NF-12-2-0023. The views and conclusions contained in this document are those of the authors and should not be interpreted as representing the official policies, either expressed or implied, of the Army Research Laboratory or the U.S. Government. The U.S. Government is authorized to reproduce and distribute reprints for Government purposes notwithstanding any copyright notation herein. 
\end{acknowledgments}

\bibliography{paperanapolerv2}

\end{document}